\documentclass[11pt,a4paper]{article}
\usepackage{jheppub}
\usepackage[latin9]{inputenc}
\usepackage{amsmath}
\usepackage{amssymb}
\usepackage{setspace}
\usepackage{graphicx}
\usepackage{babel}
\usepackage[caption=false]{subfig}

\newcommand{\tr}{\mathrm{tr}}
\newcommand{\e}{\mathrm{e}}

\allowdisplaybreaks

\usepackage{comment}

\begin{document}

\title{Fermionic domain-wall Skyrmions of QCD in a magnetic field}

\author[a]{Patrick~Copinger}
\emailAdd{copinger@hiroshima-u.ac.jp}
\affiliation[a]{International Institute for Sustainability with Knotted Chiral Meta Matter (WPI-SKCM$^2$), Hiroshima University, 1-3-2 Kagamiyama, Higashi-Hiroshima, Hiroshima 739-8511, Japan}
\author[b,c,a]{Minoru~Eto}
\emailAdd{meto@sci.kj.yamagata-u.ac.jp}
\affiliation[b]{Department of Physics, Yamagata University, 
Kojirakawa-machi 1-4-12, Yamagata, Yamagata 990-8560, Japan}
\affiliation[c]{Research and Education Center for Natural Sciences, Keio University, 4-1-1 Hiyoshi, Yokohama, Kanagawa 223-8521, Japan}
\author[c,d,a]{Muneto~Nitta}
\emailAdd{nitta@phys-h.keio.ac.jp}
\affiliation[d]{Department of Physics, Keio University, 4-1-1 Hiyoshi, Yokohama, Kanagawa 223-8521, Japan}
\author[c]{Zebin Qiu}
\emailAdd{qiuzebin@keio.jp}

\abstract{
The ground state of 
low-energy QCD matter in  strong magnetic fields is either a chiral soliton lattice 
(CSL), a periodic array of neutral pion domain walls (chiral solitons)
perpendicular to the magnetic field, or domain-wall Skyrmion phase, 
in which Skyrmions are induced on top of 
the CSL.
Previously found domain-wall Skyrmions 
are bosons with the baryon number two.
In this paper, we show that 
the minimum domain-wall Skyrmions are fermions 
with baryon number one; 
a bosonic domain-wall Skyrmion 
can be separated 
without energy cost 
into two 
fermionic domain-wall Skyrmions 
attached on the opposite sides of a chiral soliton.
The phase boundary between the CSL 
and domain-wall Skyrmion phases is unchanged. 
In the chiral limit, the CSL reduces to a 
linearly dependent neutral pion on the direction of the magnetic field, while fermionic domain-wall Skyrmions sit in an equal distance of half a period.
}

\maketitle

%%%%%%%%%%%%%%%%%%%%%%%%%%%%%%%%%%%%%%%%%%%%%%%%%%%%%%%%%

\section{Introduction}

The makeup of the phase diagram of quantum chromodynamics (QCD) is a pertinent and outstanding question in physics, for its understanding predicts the formation of nuclear matter--and hence the abundance of visible mass in the universe~\cite{Fukushima:2010bq,Schmidt:2017bjt,Fischer:2018sdj}. Further, compli\-menting our theoretical understanding of the phase diagram is the potential observability of its features in heavy-ion collision experiments, making the topic one of intense active investi\-gation: e.g., investigations of the tricritical point~\cite{STAR:2025zdq} or phase transitions~\cite{Behera:2025oyq} from the STAR collaboration at the Relativistic Heavy Ion Collider, or the conserved quantum number fluctuation at ALICE~\cite{ALICE:2025mkk}. At colliders, in addition to a large density, both a strong magnetic field in off-central collisions as well as a rapid rotation of the nuclei are present~\cite{Fukushima:2010bq}, which motivate studies of the phase diagram of QCD. In particular the presence of a density, and hence chemical potential, in QCD beget the sign problem, hampering one of our main theoretical tools of lattice QCD. To garner insight into the phase diagram in such extreme environments, chiral perturbation theory (ChPT)~\cite{Gell-Mann:1960mvl,Weinberg:1978kz,Gasser:1983yg,GASSER1985465,Leutwyler:1993iq} proves to be a valuable tool. ChPT is a low-energy theory of QCD built on its observed symmetries, most notably chiral symmetry breaking leading to the appearance of Nambu-Goldstone (NG) bosons or pions~\cite{Scherer:2012xha,Bogner:2009bt}. Incorporation of finite baryon chemical potential $\mu_B$ or strong background magnetic field is facilitated with the Wess-Zumino-Witten (WZW) term, which contains an anomalous coupling of the neutral pion to a magnetic field, predicted through the chiral anomaly~\cite{Son:2004tq,Son:2007ny}, and even the chiral separation effect~\cite{Son:2004tq,PhysRevD.22.3080,Metlitski:2005pr,Landsteiner:2016led}.

For QCD at finite density, and with a strong magnetic field, ChPT with WZW coupling predicts many interesting features. Most notably the ground state of the theory with two flavors of up and down quarks is modified as a stack of domain walls with baryon number coined the chiral soliton lattice (CSL)~\cite{Son:2007ny,Eto:2012qd,Brauner:2016pko}. CSLs in QCD have been examined with the inclusion of thermal fluctuations~\cite{Brauner:2017uiu,Brauner:2017mui,Brauner:2021sci,Brauner:2023ort}, under rotation~\cite{Huang:2017pqe,Nishimura:2020odq,Chen:2021aiq,Eto:2021gyy,Eto:2023tuu,Eto:2023rzd}, in their quantum~\cite{Eto:2022lhu,Higaki:2022gnw} and dynamical~\cite{Eto:2025ebz} nucleation, from the framework of holography~\cite{Amano:2025iwi}, in the formation of Abrikosov-Nielsen-Olesen vortices of its charged pions~\cite{Qiu:2024zpg,Hamada:2025inf}, an incommensurate lattice made of neutral pion and $\eta'$ solitons~\cite{Qiu:2023guy} among over studies~\cite{Yamada:2021jhy,Brauner:2019aid, Brauner:2019rjg,Nitta:2024xcu,Canfora:2025aau}. However, the CSL ground state is an unstable one: One of which occurs for high density and or strong magnetic fields in which a tachyon and a charged pion condensation (CPC) appear; this is characterized at an asymptotic magnetic field strength of $B_{\text{CPC}}\sim16\pi^4f_\pi^4/\mu_B^2$~\cite{Brauner:2021sci}. Beyond CPC, baryon crystals have been proposed to appear~\cite{Evans:2022hwr,Evans:2023hms}. Another instability that occurs for magnetic field strengths less than $B_{\text{CPC}}$ and is characterized with the appearance of domain-wall Skyrmions (DWSk). DWSks, or the combined state of a domain wall and a Skyrmion, have had a rich history: They were first studied in quantum field theory (QFT) environments in (2+1)-dimensions in refs.~\cite{Nitta:2012xq,Kobayashi:2013ju}, and then studied in (3+1)-dimensions in refs.~\cite{Nitta:2012wi,Nitta:2012rq,Gudnason:2014hsa,Gudnason:2014nba,Eto:2015uqa}. DWSks have also been analyzed in a (2+1)-dimensional variant in condensed matter~\cite{PhysRevB.99.184412,Kuchkin:2020bkg,Ross:2022vsa,Amari:2023gqv,Amari:2023bmx,Amari:2024jxx,Gudnason:2024shv,Leask:2024dlo,Lee:2024lge}, and even observed in a chiral magnet in refs.~\cite{Nagase:2020imn,Yang:2021nem}. In QCD under a strong magnetic field DWSks have been studied beyond the BPS approximation to incorporate gauge field dynamics~\cite{Amari:2024fbo} and rapid rotation, as is important in heavy-ion collider and neutron star environments~\cite{Eto:2023tuu}

In previous works on DWSks in QCD with a magnetic field, import\-antly it was found that one lump in the soliton would correspond to two Skyrmions in the bulk, and thus with an even baryon number~\cite{Eto:2025fkt,Eto:2023wul}, with moreover \textit{boson} statistics~\cite{Amari:2024mip}. A priori it is unclear whether the constituents of the DWSk theory separately predict ground state and or phase diagram features as predicted in the combined theory, and for that matter whether the constituents are even \textit{fermionic} at all. In this paper we rectify this issue, and guided by intuition\footnote{We can gather further intuition from the case of the DWSk in QCD under rapid rotation in which one lump on the soliton is a single Skyrmion~\cite{Eto:2023tuu}.} provided from the macaron-like profile of the Skyrmion charge density~\cite{Eto:2025fkt,Eto:2023wul}, we construct `half' effective DWSk theories whose Skyrmions are quantized in integer number, and whose statistics we find are indeed fermionic. Thus the minimum DWSk is a fermion with baryon number one. To further support our findings we promote the lump's position moduli parameter considering the energy of an effective theory, and conclude the ground state energy of the half DWSk effective theory is saturated and a robust feature of the theory.

The organization of this paper is as follows: In sec.~\ref{sec:csl} we review the effective chiral perturbative Lagrangian under a magnetic field, which leads to the CSL. In sec.~\ref{sec:half} we construct the half period chiral solitonic theory leading to Skyrmions in the bulk with fermionic statistics. And in sec.~\ref{sec:separation} we look at their separation energy under the position moduli parameter. In sec.~\ref{sec:chiral} we discuss the chiral limit of our effective theory. And last in sec.~\ref{sec:conclusions} we offer closing discussions and some future directions.

%%%%%%%%%%%%%%%%%%%%%%%%%%%%%%%%%%%%%%%%%%%%%%%%%%%%%%%%%
\section{QCD chiral soliton lattice with magnetic field}
\label{sec:csl}

We take as our starting point the chiral Lagrangian of ChPT for $N_F=2$ flavors in a background electromagnetic magnetic field with the WZW coupling~\cite{Wess:1971yu,Witten:1983tx,Son:2007ny} to account for the chiral anomaly. In this section we briefly review the theory's notations and CSL  setup from the magnetic field.

%%%%%%%%%%%%%%%%%%%%%%%%%%%%%%%%%%%%%%%%%%%%%%%%%%%%%%%%%
\subsection{Effective chiral perturbative Lagrangian}

We study the effective ChPT Lagrangian, keeping terms to $\mathcal{O}(p^2)$~\cite{Brauner:2021sci,Eto:2023wul}, with a mostly minus signature for (3+1)-dimensions:
\begin{equation}
    \mathcal{L}_{\textrm{ChPT}}=\frac{f_\pi^2}{4}\tr \bigl(D_\mu\Sigma  D^\mu\Sigma \bigr)
    -\frac{f_\pi^2m_\pi^2}{4}\bigl(2-\Sigma - \Sigma^\dagger  \bigr)\,.
\end{equation}
$f_\pi$ and $m_\pi$ are respectively the pion decay constant and mass. The pion d.o.f. are contained in the SU$(2)$ weighted pion field
\begin{equation}
    \Sigma = \exp(i\tau_a\chi_a)=\frac{1}{f_\pi}(\sigma +i\tau^a\pi^a)\,,
\end{equation}
for Pauli matrix generator, $\tau_a$. The matrices are normalized such that $\tr(\tau_a \tau_b)=2\delta_{ab}$. The charged pions are constructed as $\pi^\pm =\pi^1\mp i\pi^2$, and the pions fulfill the relation $\sigma^2 + \pi^a\pi^a=f_\pi^2$. The covariant derivative reads
\begin{equation}
    D_\mu \Sigma = \partial_\mu \Sigma +ieA_\mu[Q,\Sigma]\,,\qquad Q=\frac{\tau^3}{2}+\frac{1}{6}\,.
\end{equation}
Symmetries present in the theory include: the pion field transformation under SU$(2)_L\times$SU$(2)_R$ as $\Sigma\to L\Sigma R^\dagger$, and the electromagnetic U$(1)$ gauge transformation:
\begin{equation}
    \Sigma\to\e^{iQ\alpha(x)}\Sigma\e^{-iQ\alpha(x)}\,,\quad
    \pi^\pm\to\e^{\pm i\alpha(x)}\pi^\pm\,,\quad
    A_\mu\to A_\mu -\frac{1}{e}\partial_\mu \alpha(x)\,.
\end{equation}

Accompanied to the ChPT Lagrangian is the Lagrangian associated with the WZW term to correctly account for chiral anomalous effects. The WZW term consists of a coupling of the U(1)$_{B}$ gauge field to the pions via the Goldstone-Wilczek current~\cite{PhysRevLett.47.986,Witten:1983tx,Son:2007ny}
\begin{equation}
    \mathcal{L}_{\textrm{WZW}}=-A_{\mu}^{B}j_{\textrm{GW}}^{\mu}=A_{\mu}^{B}\frac{\epsilon^{\mu\nu\alpha\beta}}{24\pi^{2}}\mathrm{tr}\bigl(L_{\nu}L_{\alpha}L_{\beta}-3ie\partial_{\nu}[A_{\alpha}Q(L_{\beta}+R_{\beta})]\bigr)\,,
    \label{eq:WZW_L}
\end{equation}
where $L_\mu=\Sigma\partial_\mu\Sigma^\dagger$ and $R_\mu=(\partial_\mu\Sigma^\dagger)\Sigma$. We fix the totally anti-symmetric tensor so that $\epsilon^{0123}=-\epsilon_{0123}=1$. With the WZW term our total effective Lagrangian is
\begin{equation}
    \mathcal{L}=\mathcal{L}_\textrm{ChPT} + \mathcal{L}_\textrm{WZW}\,.
    \label{eq:full_L}
\end{equation}

Last let us turn to the background gauge couplings. For the U$(1)_\textrm{EM}$ electromagnetic coupling, for simplicity, our strong magnetic field lies in the $\hat{x}^3$ direction, $\boldsymbol{B}=B\hat{x}^3$ with gauge $A_\mu=(B/2)(0,x^2,-x^1,0)$. We also treat, through the U$(1)_B$ baryonic gauge field, a non-zero baryon chemical potential given by $A_\mu^B=(\mu_B,0,0,0)$. In the presence of the strong magnetic field, a CSL  develops as the ground state.

%%%%%%%%%%%%%%%%%%%%%%%%%%%%%%%%%%%%%%%%%%%%%%%%%%%%%%%%%
\subsection{Chiral soliton lattice (full period)}

To show the dynamics of the CSL we need only analyze $\chi_3$, ignoring the charged pions, with
\begin{equation}
    \Sigma_0=\exp(i\tau_3\chi_3)\,.
\end{equation}
A charged pion contribution can be incorporated by performing an SU$(2)_V$ transformation, $\Sigma=g^\dagger\Sigma g$ with $g\in$ SU$(2)_V$. The above reduction is sufficient to show the presence of sine-Gordon soliton solutions in a strong magnetic field~\cite{Son:2007ny}. Then inserting $\Sigma\to\Sigma_0$ into eq.~\eqref{eq:full_L} the ChPT effective Lagrangian becomes
\begin{equation}
    \mathcal{L}_\textrm{ChPT}=-\frac{f_\pi}{2}(\partial_3\chi_3)^2
    -f_\pi^2 m_\pi^2 (1-\cos\chi_3)\,,
\end{equation}
and the WZW contribution is $\mathcal{L}_\textrm{WZW}=(eB\mu_B/4\pi^2) \partial_3\chi_3$. The resulting equation of motion, unaffected by the boundary WZW term, is
\begin{equation}
    \partial_3^2\chi_3^\textrm{CSL}-m^2_\pi\sin\chi_3^\textrm{CSL}=0\,.
\end{equation}
with sine-Gordon soliton solution
\begin{equation}
    \chi_3^\textrm{CSL}=2\mathrm{am}\Bigl(\frac{m_\pi z}{\kappa},\kappa\Bigr) + \pi\,,
    \label{eq:CSL_solution}
\end{equation}
with elliptic modulus, $\kappa$, where $0\leq\kappa \leq 1$, which dictates the density of stacked lattices in the CSL. Here $x^3=z$. The neutral pion solution satisfies $\chi_3^\textrm{CSL}(z+\ell)=\chi_3^\textrm{CSL}(z) +2 \pi$ with period
\begin{equation}
    \ell=\frac{2\kappa K(\kappa)}{m_\pi}\,,
\end{equation}
where $K(\kappa)$ is the complete elliptic integral of the first kind. 

Finally let us review the energy density per unit area, or tension over one \textit{full} period for a single soliton. This is given as $\sigma_\textrm{CSL}=\int^\ell_0dz\,\mathcal{H}$, for Hamiltonian of the effective Lagrangian, $\mathcal{L}$, evaluated on a single soliton solution, eq.~\eqref{eq:CSL_solution}. For the \textit{full} entire period effective energy per unit area we have that
\begin{equation}
    \sigma_\textrm{CSL} = 4m_\pi f_\pi^2 \Bigl[ \frac{2E(\kappa)}{\kappa} +\Bigl( \kappa-\frac{1}{\kappa} \Bigr) K(\kappa)\Bigr]-\frac{eB\mu_B}{2\pi}\,.
\end{equation}
When $\sigma_\textrm{CSL}<0$ this indicates that the sine-Gordon soliton is more energetically favorable than is the QCD vacuum. Minimizing the tension over $\kappa$ a bound on the magnetic field and chemical potential can be found as
\begin{equation}
    \frac{E(\kappa)}{\kappa}=\frac{eB\mu_B}{16\pi f_\pi^2m_\pi}\,.
    \label{eq:kappa_constraint}
\end{equation}

\begin{figure}
\centering
\includegraphics[scale=0.8]{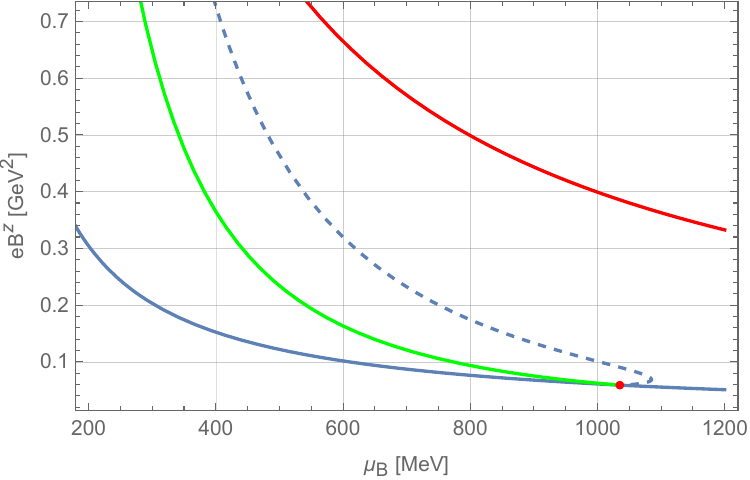}
\caption{The QCD phase diagram of baryon chemical potential and magnetic field. Below the solid blue line lies the QCD vacuum~\cite{Son:2007ny}. In between the green and blue lines lies the CSL phase. Above the green line lies the domain wall Skyrmion phase. And above the dotted blue line, the CSL becomes unstable depicted by the charged pion condensation above $B~\sim16\pi^4f^4_\pi / \mu_B^2$~\cite{Brauner:2016pko}. The red line describes the critical region about $B_\text{max}$ where ChPT is no longer valid. }
\label{fig:phase}
\end{figure}

The magnetic field and baryon chemical potential in question predicting the CSL, and as we will discover below Skyrmions embedded in the domain wall, beg the question of applicability of ChPT, and to what extend are our theories valid. This problem has been studied in Ref.~\cite{Brauner:2016pko}, and we further show a $B_\text{max}=|\vec{B}_\text{max}|$ that illustrates a critical magnetic field where ChPT do longer applies. ChPT is facilitated through a derivative expansion about the parameter $p/(4\pi f_\pi)$ with characteristic momentum, $p$~\cite{Brauner:2016pko}. Hence one must have that $p\ll 4\pi f_\pi$. The CSL is characterized by the field $\chi^\textrm{CSL}_3$ given in~\eqref{eq:CSL_solution}, and to determine the limiting bound we need only examine the chiral limit form, whose expression is given as $\chi^\textrm{CSL}_3=e\mu_BBz/(4\pi^2f_\pi^2)$ up to an irrelevant constant. This field defines our momentum scale as $p_{\text{CSL}}=e\mu_BB/(4\pi^2f_\pi^2)$. Then considering that we must have $p\ll 4\pi f_\pi$, we can determine the bounding magnetic field of our theory as 
\begin{equation}
    B_\text{max}=\frac{16\pi^3 f_\pi^3}{\mu_B}\,,
\end{equation}
where we must have that $B<B_\text{max}$. This lies well-above the values considered here and thus our treatment of ChPT is indeed valid. See Fig.~\ref{fig:phase} for details of the QCD phase diagram in a baryon chemical potential and magnetic field. 

%%%%%%%%%%%%%%%%%%%%%%%%%%%%%%%%%%%%%%%%%%%%%%%%%%%%%%%%%
\section{Effective solitonic theory over half period in CSL}
\label{sec:half}

Having explored some of the basic known setup, let us now consider the half domain wall Skyrmion and the baryon(s) that live on it. As seen in ref.~\cite{Eto:2023wul}, the effective theory given over the entire period of the DWSk furnishes quantized topological charges, through the WZW term, corresponding to baryon number two, with bosonic statistics. We confirm here that the half effective theory indeed can correspond to a single baryon, with fermionic statistics. Let us first consider the effective ChPT theory with charged pions.

\subsection{Effective chiral perturbative Lagrangian}

We consider the effective theory of a non-Abelian sine-Gordon soliton using the known moduli approximation~\cite{MANTON198254,Eto:2006uw}. First we turn on the electric charge of the pions considering the SU$(2)_V$ transformation of the CSL as
\begin{equation}
    \Sigma=g\Sigma_0g^\dagger =\exp(i\chi_3^{\textrm{CSL}}g\tau_3g^\dagger)\,.
    \label{eq:Sig_def}
\end{equation}
This is a nontopological non-Abelian sine-Gordon soliton, in contrast to topological counter\-part in refs.~\cite{Nitta:2014rxa,Eto:2015uqa}, whose moduli are in the coset space of $\mathcal{M}\cong\text{SU}(2)_V/\text{U}(1)_3\cong \mathbb{C}P^1$, owing to the redundancy of the U$(1)_3$ subgroup in $g$ from $\tau_3$. Let us parameterize the moduli with coordinates $\phi\in\mathbb{C}^2$ that obey
\begin{equation}
    \phi^\dagger \phi =1\,\quad g\tau_3g^\dagger=2\phi\phi^\dagger-\boldsymbol{I}_2\,.
\end{equation}
Under the new coordinates we may write
\begin{equation}
    \Sigma=[\boldsymbol{I}_2+(u^2-1)\phi\phi^\dagger]u^{-1}\,,
    \label{eq:Sigma}
\end{equation}
where we have $u=\exp(i\chi_3^\textrm{CSL})$ where $\chi_3^\textrm{CSL}$ is given by eq.~\eqref{eq:CSL_solution}.

Under the moduli approximation the effective chiral perturbative Lagrangian becomes~\cite{Eto:2023wul}
\begin{align}
    \mathcal{L}_{\textrm{ChPT}} & =\frac{f_{\pi}^{2}}{2}|1-u^{2}|^{2}\bigl[\bigl((\phi^{\dagger}\partial_{\alpha}\phi)^{2}+\partial_{\alpha}\phi^{\dagger}\partial^{\alpha}\phi\bigr)\bigr]\notag\\
    &+i\frac{e}{2}f_{\pi}^{2}A^{\alpha}|1-u^{2}|^{2}\Bigl[\phi^{\dagger}\tau_{3}\phi\cdot\phi^{\dagger}\partial_{\alpha}\phi+\frac{1}{2}\bigl(\partial_{\alpha}\phi^{\dagger}\tau_{3}\phi-\phi^{\dagger}\tau_{3}\partial_{\alpha}\phi\bigr)\Bigr]\nonumber \\
    &-\frac{e^{2}f_{\pi}^{2}}{8}A^{2}|1-u^{2}|^{2}\bigl[(\phi^{\dagger}\tau_{3}\phi)^{2}-1\bigr]-\mathcal{H}_\textrm{ChPT}\,,
\end{align}
with corresponding Hamiltonian
\begin{equation}
        \mathcal{H}_\textrm{ChPT} =\frac{f_{\pi}^{2}}{2}(\partial_{z}\chi_{3}^{\textrm{CSL}})^{2}+f_{\pi}^{2}m_{\pi}^{2}(1-\cos\chi_{3}^{\textrm{CSL}})\,.
\end{equation}
One may further reduce the kinetic contributions in the effective Lagrangian by organizing terms into a covariant derivative with $D_{\mu}=\partial_{\mu}+ieA_{\mu}\tau_{3}/2$; the Lagrangian then becomes
\begin{equation}
    \mathcal{L}_{\textrm{ChPT}}=\frac{f_{\pi}^{2}}{2}|1-u^{2}|^{2}\bigl[\bigl((\phi^{\dagger}D_{\alpha}\phi)^{2}+(D_{\alpha}\phi)^{\dagger}D^{\alpha}\phi\bigr)\bigr]-\mathcal{H}_\textrm{ChPT}\,.
\end{equation}
We remark at this point that in addition to the $\mathbb{C}P^1$ moduli for single soliton solution the full moduli is $\mathcal{M}\cong \mathbb{R}\times\mathbb{C}P^1$ that encompasses the translation modulus of $z\to z+Z$ where $Z$ may also be extended to live on a (2+1)-d world volume. With the additional translation modulus term the effective Hamiltonian is $\mathcal{H}\to \mathcal{H}-(f_{\pi}^{2}/2)\partial_{\mu}Z\partial^{\mu}Z(\partial_{z}\chi_{3}^{\textrm{CSL}})^{2}$. Such a distinction; however, will not alter the phase diagram structure of our half theory. Therefore we treat here and throughout the moduli approximation of $\mathcal{M}\cong \mathbb{C}P^1$. 

We are now in a position to determine the components of the effective domain wall Lagrangian on the half period. For simplicity here and throughout we will treat the half from 0 to $\ell/2=\kappa K(\kappa)/m_{\pi}$. Let us first determine the kinetic term. Using the fact that 
\begin{align}
    \int_{0}^{\ell/2}dz\,|1-u^{2}|^{2}& =16\int_{0}^{\kappa K(\kappa)/m_{\pi}}dz\,\Bigl[\mathrm{sn}\Bigl(\frac{m_{\pi}z}{\kappa},\kappa\Bigr)\Bigr]^{2}\Bigl[\mathrm{cn}\Bigl(\frac{m_{\pi}z}{\kappa},\kappa\Bigr)\Bigr]^{2} \nonumber\\
    & =\frac{16}{3m_{\pi}}\frac{1}{\kappa^{3}}\Bigl[(2-\kappa^{2})E(\kappa)+2(\kappa^{2}-1)K(\kappa)\Bigr]\,,
\end{align}
where $K(\kappa)$ and $E(\kappa)$ are respectively the complete elliptic integral of first and second kind. Thus the kinetic term here becomes one half of that of the full effective theory's term:
\begin{equation}
    \frac{f_{\pi}^{2}}{2}\int_{0}^{\ell/2}dz\,|1-u^{2}|^{2}\bigl[(\phi^{\dagger}D_{\alpha}\phi)^{2}+(D_{\alpha}\phi)^{\dagger}D^{\alpha}\phi\bigr] =\mathcal{C}(\kappa)\frac{1}{2}\bigl[(\phi^{\dagger}D_{\alpha}\phi)^{2}+(D_{\alpha}\phi)^{\dagger}D^{\alpha}\phi\bigr]\,,
\end{equation}
with K\"ahler class given by
\begin{equation}
    \mathcal{C}(\kappa) =\frac{16f_{\pi}^{2}}{3m_{\pi}}\frac{1}{\kappa^{3}}\Bigl[(2-\kappa^{2})E(\kappa)+2(\kappa^{2}-1)K(\kappa)\Bigr]\,.
\end{equation}

Let us next treat the half effective part of the Hamiltonian associated with the CSL without charged pions, and further without the WZW term--as that will play a prominent role later. Again we define the half effective theory as one after integrating out the half period from  0 to $\ell/2$.
\begin{align}
    \mathcal{E}_{\tfrac{1}{2}\textrm{ChPT}} & =\int_{0}^{\ell/2}dz\,\mathcal{H}_\textrm{ChPT}
    \nonumber\\
    & =2f_{\pi}^{2}\frac{m_{\pi}}{\kappa}\int_{0}^{K(\kappa)}d\bar{z}\,[\mathrm{dn}(\bar{z},\kappa)]^{2}+2f_{\pi}^{2}m_{\pi}\kappa K(\kappa)-2f_{\pi}^{2}m_{\pi}\kappa\int_{0}^{K(\kappa)}dz[\mathrm{sn}(\bar{z},\kappa)]^{2}
    \nonumber\\
    & =2f_{\pi}^{2}m_{\pi}\Bigl[\frac{2}{\kappa}E(\kappa)+\Bigl(\kappa-\frac{1}{\kappa}\Bigr)K(\kappa)\Bigr]\,.
    \label{eq:E1/2}
\end{align}
Similar to the tension of one full period of the CSL, we find here as well, after minimizing w.r.t. the elliptic modulus, $\kappa$, that eq.~\eqref{eq:kappa_constraint} is still valid in the half theory. We also remark that incorporating the translation modulus would augment the above by 
\begin{equation}
    \mathcal{E}_{\tfrac{1}{2}\textrm{ChPT}}\to \mathcal{E}_{\tfrac{1}{2}\textrm{ChPT}}-2f_{\pi}^{2}\frac{m_{\pi}}{\kappa}E(\kappa)\partial_{\mu}Z\partial^{\mu}Z\,,
\end{equation}
but otherwise would not change qualitatively the phase diagram.

\subsection{WZW contribution}

Having determined the kinetic contributions and Hamiltonian to our effective half theory, let us now turn our attention to the WZW term given in eq.~\eqref{eq:WZW_L}. We decompose the WZW term into two parts: the Skyrmion charge and the electromagnetic coupling term. Let us first treat the former.

\begin{figure}
\centering
\subfloat[]{%
\includegraphics[scale=0.5]{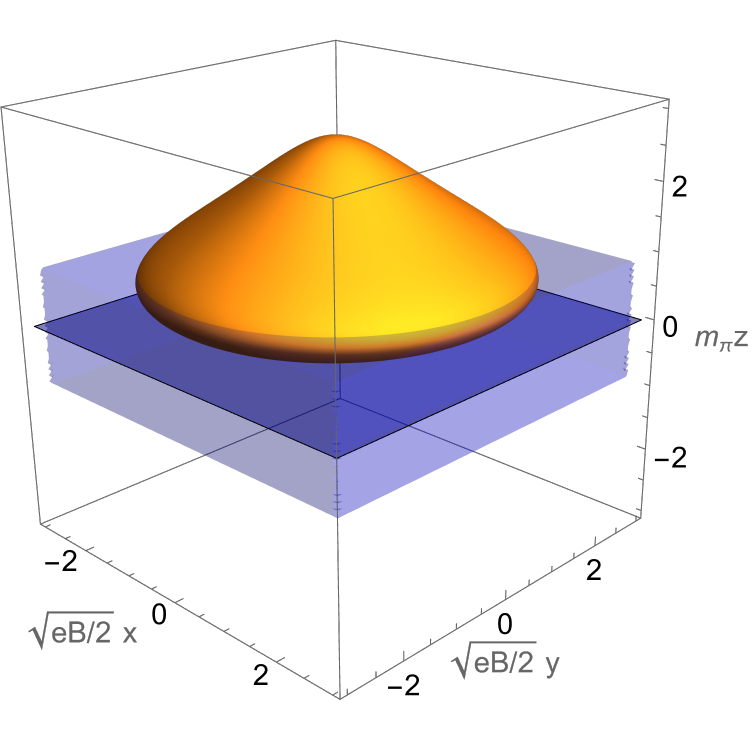}}
\subfloat[]{%
\includegraphics[scale=0.5]{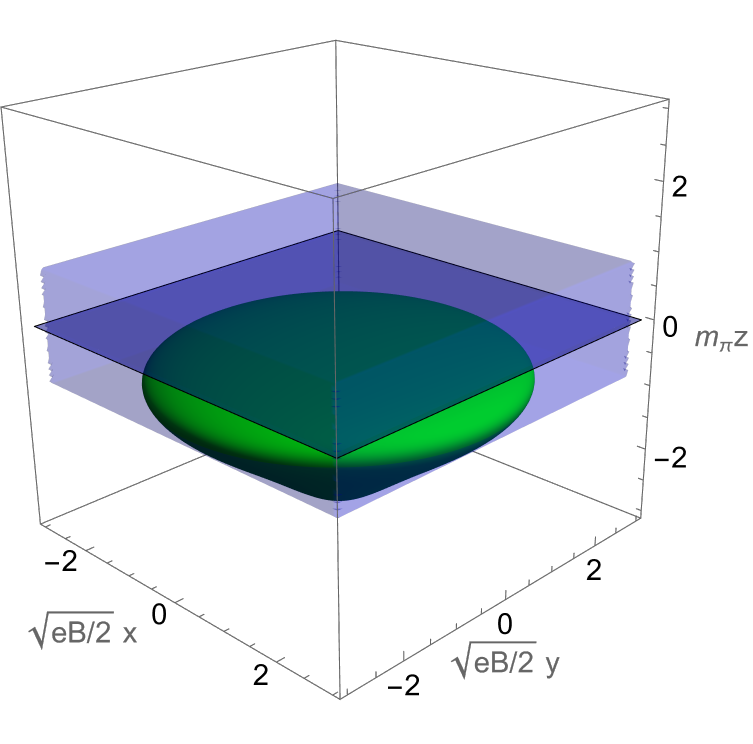}}
\caption{
Isosurfaces of $\mathcal{B}=1/(50\pi^2)$ for both effective DWSk theories on (a) $0<z<\ell/2$ shown in amber and (b) $\ell/2<z<\ell$ shown in green. Both are for the single soliton case, $\kappa=1$. The CSL in blue extending into both halves of (a) and (b) is shown for $\pi/2<\chi_3^{\textrm{CSL}}<3\pi/2$ with blue plane intersecting each half. Notice the separation of each macaron across each intersecting plane depicting the separation of baryon topological charges $k=1$ and $k=1$ for each half. 
}
\label{fig:1_half_macaron}
\end{figure}

Let us last calculate the Skyrmion charge density from a half period following from 
\begin{equation}
    \mathcal{B}=\frac{-1}{24\pi^{2}}\epsilon^{ijk}\mathrm{tr}[L_{i}L_{j}L_{k}]\,.
\end{equation}
Its solution under the moduli approximation employed above reads
\begin{equation}
    \mathcal{B}=-\frac{1}{2\pi}\Bigl(u-\frac{1}{u}\Bigr)^{2}\partial_{z}\chi_{3}^{\text{CSL}}q(x,y)\,.\label{eq:beta}
\end{equation}
Here $q(x,y)$ is the $\mathbb{C}P^{1}$ lump topological charge density
such that 
\begin{equation}
    \int dxdy\,q(x,y)=k\in\pi_{2}(\mathbb{C}P^{1})\,.
\end{equation} 
Now similar to as was accomplished for the kinetic and Hamiltonian terms, one may readily find the half period DWSk effective term by integrating $\mathcal{B}$ over a half period; this results in 
\begin{equation}
    \int_{0}^{\kappa K(\kappa)/m_{\pi}}dz\,\mathcal{B}=q(x,y)\,,
    \label{eq:q}
\end{equation}
where we have made use of
\begin{equation}
    -\frac{1}{2\pi}\int_{0}^{\kappa K(\kappa)/m_{\pi}}dz\,\Bigl(u-\frac{1}{u}\Bigr)^{2}\partial_{z}\chi_{3}^{\text{CSL}} =\frac{16}{\pi}\int_{0}^{K(\kappa)}d\bar{z}\,\mathrm{sn}(\bar{z},\kappa)^{2}\mathrm{cn}(\bar{z},\kappa)^{2}\mathrm{dn}(\bar{z},\kappa)=1\,.
\end{equation}
As anticipated, we have found that the effective half theory represents one of a \textit{single} baryon (for Skyrmion topological winding number $\int dxdy \,q(x,y)=1$). We emphasize that the theory with the full period, i.e., from $0$ to $\ell=2 \kappa K(\kappa)/m_{\pi}$, produces $\int^\ell_0dz\,\mathcal{B}=2q(x,y)$ and a theory with \textit{two} baryons--with moreover bosonic statistics (also for Skyrmion topological winding number $\int dxdy \,q(x,y)=1$). Also by the natural extension from both integrals we can indeed confirm that the other half is 
\begin{equation}
    \int^{2\kappa K(\kappa)/m_{\pi}}_{\kappa K(\kappa)/m_{\pi}}dz\,\mathcal{B}=q(x,y)\,,
\end{equation}
also in support of the finding that the half period is represented with at least single baryon number. Let us show the structure of the separated Skyrmions in fig.~\ref{fig:1_half_macaron} where we show an isosurface plot for both halves with baby Skyrmion charge density evaluated to 
\begin{equation}
    q(x,y)=\frac{1}{\pi}\frac{2}{eB}\frac{1}{(x^{2}+y^{2}+\frac{2}{eB})^{2}}\,,
    \label{eq:baby_Skyrmion}
\end{equation}
for the single lump with $\omega=x+iy$.
We will shortly prove the fermionic statistics for half theory. However, let us first continue with the evaluation of the rest of the WZW term in the DWSk effective theory. 

We now compute the right hand side of eq.~\eqref{eq:WZW_L} with the electromagnetic coupling. We follow along the steps in ref.~\cite{Eto:2023wul} to arrive at the following breakdown of terms:
\begin{align}
 & \frac{ie\mu_{B}\epsilon^{0\nu\alpha\beta}}{16\pi^{2}}\partial_{\nu}[A_{\alpha}\mathrm{tr}[\tau_{3}(L_{\beta}+R_{\beta})]]\nonumber \\
 & =-\frac{e\mu_{B}B}{4\pi^{2}}(\phi^{\dagger}\tau_{3}\phi)\partial_{3}\chi_{3}^{\textrm{CSL}}-\frac{e\mu_{B}}{8\pi^{2}}\epsilon^{03jk}|1-u^{2}|^{2}\partial_{3}\chi_{3}^{\textrm{CSL}}A_{j}\partial_{k}(\phi^{\dagger}\tau_{3}\phi)\,.
\end{align}
And then let us go ahead and integrate out a half period to find the effective term becomes
\begin{align}
    & \frac{ie\mu_{B}\epsilon^{0\nu\alpha\beta}}{16\pi^{2}}\int_{0}^{\ell/2}dz\,\partial_{\nu}[A_{\alpha}\mathrm{tr}[\tau_{3}(L_{\beta}+R_{\beta})]] \nonumber\\
    & =-\frac{e\mu_{B}B}{2\pi^{2}}(\phi^{\dagger}\tau_{3}\phi)\int_{0}^{K(\kappa)}d\bar{z}\,\mathrm{dn}(\bar{z},\kappa)\nonumber \\
    & \quad-\frac{4e\mu_{B}}{\pi^{2}}\epsilon^{0ijk}A_{j}\partial_{k}(\phi^{\dagger}\tau_{3}\phi)\int_{0}^{K(\kappa)}d\bar{z}\,[\mathrm{sn}(\bar{z},\kappa)]^{2}[\mathrm{cn}(\bar{z},\kappa)]^{2}\mathrm{dn}(\bar{z},\kappa)\nonumber\\
    & =-\frac{e\mu_{B}}{4\pi}\epsilon^{03jk}\partial_{j}(A_{k}\phi^{\dagger}\tau_{3}\phi)\,.
    \label{eq:wzw_final}
\end{align}
Indeed half of what the term is for the full DWSk theory as expected.

Finally, gathering everything together from eqs.~\eqref{eq:E1/2},~\eqref{eq:q} and~\eqref{eq:wzw_final} we find for the entire half theory effective domain wall Lagrangian the following:
\begin{align}
\mathcal{L}_{\textrm{DW}} & =-2f_{\pi}^{2}m_{\pi}\Bigl[2\frac{1}{\kappa}E(\kappa)+(\kappa-\frac{1}{\kappa})K(\kappa)\Bigr]\nonumber \\
 & \quad+\mathcal{C}(\kappa)\frac{1}{2}\bigl[(\phi^{\dagger}D_{\alpha}\phi)^{2}+(D_{\alpha}\phi)^{\dagger}D^{\alpha}\phi\bigr]-\mu_{B}q+\frac{e\mu_{B}}{4\pi}\epsilon^{03jk}\partial_{j}[A_{k}(1-n_{3})]\,,\label{eq:eff_Lag}
\end{align}
which is indeed one-half of the full-period theory. And the other half from $\ell/2$ to $\ell$ is identical to the above.

\subsection{Spin statistics}

Here we examine the spin statistics of the half domain wall Skyrmion. We employ the Witten method~\cite{Witten:1983tx} as used for various topological solitons in chiral perturbation theory with WZW and a magnetic field in ref.~\cite{Amari:2024mip}. Our setup is identical for the most part with the full theory of the domain wall Skyrmions, so we refer to ref.~\cite{Amari:2024mip} for details. 

Essentially the Witten method uses an embedding, i.e. SU$(2)$ valued pion field into a SU$(3)$ valued one, and mapping a spatial rotation in SU$(2)$ to a homotopically trivial transformation in SU$(3)$. Important to this approach is the evaluation of the 5-dimensional WZW term that sits in the effective action, and is a measure of the non-trivial response. The 5-dimensional WZW term reads
\begin{equation}
    \Gamma_{5}=\frac{iN_{C}}{240\pi^{2}}\varepsilon^{\mu\nu\alpha\beta\gamma}\int d^{5}x\,\mathrm{tr}(\tilde{L}_{\mu}\tilde{L}_{\nu}\tilde{L}_{\alpha}\tilde{L}_{\beta}\tilde{L}_{\gamma})\,,
    \label{eq:Gamma5}
\end{equation}
where $\tilde{L}=\tilde{U}(\boldsymbol{x},t,\xi)\partial_\mu\tilde{U}^\dagger(\boldsymbol{x},t,\xi)$, and $\tilde{U}(x,t,\xi)$ is the 5-dimensional extension of 
\begin{equation}
    U(\boldsymbol{x})=\begin{pmatrix} \Sigma(\boldsymbol{x})&\\
    &1 \end{pmatrix}\in \text{SU}(3)\,,
\end{equation}
for pion field given in ref.~\eqref{eq:Sig_def} such that $\tilde{U}(\boldsymbol{x},t,\xi)=V(t,\xi) U(\boldsymbol{x})V^{-1}(t,\xi)$, with 5-dimensional transformation matrix
\begin{equation}
    V(t,\xi)=\begin{pmatrix}
        1&&\\ &\xi e^{-it}&-\sqrt{1-\xi^2}\\&\sqrt{1-\xi^2}&\xi e^{it}
    \end{pmatrix}\,.
\end{equation}
The essential measurement of $\Gamma_5$ is as a response under $2\pi$ rotations in space pointing to the spin statistics of the pion field, i.e., for $\Gamma_5= 2\pi N_C k$ or $\Gamma_5= \pi N_C k$ with $k\in\mathbb{Z}$ the spin-statistics are respectively bosonic or fermionic.

The response term, eq.~\eqref{eq:Gamma5}, has been evaluated in ref.~\cite{Amari:2024mip} for the DWSk in the full theory up to the spatial integrals, and for the half theory concerned here one need only integrate along the half angle in $z$ to find the response for the half theory as
\begin{align}
\Gamma_{5} & =\frac{2\cdot3!\cdot N_{C}}{24\pi}\int_{\mathbb{R}_{3}}d^{3}\boldsymbol{x}\sin\chi\cos\chi\partial_{\rho}\chi\partial_{\varphi}\phi\partial_{z}\theta\nonumber\\
 & =\frac{2\cdot3!\cdot N_{C}}{24\pi}\int_{0}^{\infty}d\rho\partial_{\rho}\chi\int_{-\infty}^{0}dz\partial_{z}\theta\int_{0}^{2\pi}d\varphi\partial_{\varphi}\phi\sin\chi\cos\chi
 \nonumber
 \\
 & =\frac{2\cdot3!\cdot N_{C}}{24\pi}2\int_{0}^{\pi/2}d\chi\int_{-\infty}^{0}dz\partial_{z}\theta\sin\chi\cos\chi\nonumber\\
 & =\pi\nu N_{C}\,.
\end{align}
And thus we find the supporting conclusion that the half domain wall
Skyrmion obeys fermion (Boson) statistics for odd (even) winding number. Let us last confirm the DWSk phase structure of the half theory.

%%%%%%%%%%%%%%%%%%%%%%%%%%%%%%%%%%%%%%%%%%%%%%%%%%%%%%%%%
\subsection{Domain wall Skyrmion phase}
\label{sec:phase}

First let us set the covariant derivatives in the kinetic term to
$D\to\partial$, then later we will introduce the coupling. For the
above effective domain wall Lagrangian we find a corresponding Hamiltonian:
\begin{align}
\mathcal{H}_{\textrm{DW}} & =\mathcal{C}(\kappa)\frac{1}{8}[\partial_{i}\boldsymbol{n}\cdot\partial_{i}\boldsymbol{n}]+\mu_{B}q-\frac{e\mu_{B}}{4\pi}\epsilon^{03jk}\partial_{j}[A_{k}(1-n_{3})]\,.
\end{align}
And the energy that is saturated by the Bogomol'nyi bound follows
as
\begin{equation}
E_{\textrm{DW}}=\pi\mathcal{C}(\kappa)|k|+\mu_{B}k-\frac{e\mu_{B}}{4\pi}\int d^{2}x\epsilon^{03jk}\partial_{j}[A_{k}(1-n_{3})]\,.
\end{equation}

Lump solutions that satisfy the saturation are the (anti-)BPS lump
solutions. These are the same as recorded in ref.~\cite{Eto:2023wul},
and read

\begin{equation}
n_{3}=\frac{1-|f|^{2}}{1+|f|^{2}}\,,\quad f=\frac{b_{k-1}\omega^{k-1}+...+b_{0}}{\omega^{k}+a_{k-1}\omega^{k-1}+...+a_{0}}\,.
\end{equation}
Because the baryon number density is split in the half effective theory, one can identify the lump solution number, $k$, with Skymion number
and hence baryon number sitting in a solition. Further the magnetic
field dependent part of the WZW term under the BPS solution follows
as before \cite{Eto:2023wul,Eto:2025fkt}. And hence the saturated
effective domain wall energy of the half (lower in $z$) Skyrmion
is
\begin{equation}
E_{\textrm{DW}}=\pi\mathcal{C}(\kappa)|k|+\mu_{B}k+\frac{e}{2}\mu_{B}B|b_{k-1}|^{2}\,.\label{eq:energy}
\end{equation}

We can see that anti-lump solutions with $k<0$ lead to negative energy
solutions, and taking for the minimal solutions that $b_{k-1}=0$,
we find that the phase boundary in which Skyrmions are spontaneously
created is the same as the full (with both halves) theory, and is
\begin{equation}
\mu_{c}=\pi\mathcal{C}(\kappa)\,.
\label{eq:phase}
\end{equation}
The critical chemical potential would be the same as for the other
half as well. Let us next consider the separation energy between the DWSks on either side of the CSL to ensure the above feature of fermionic DWSk phase is unchanged.

%%%%%%%%%%%%%%%%%%%%%%%%%%%%%%%%%%%%%%%%%%%%%%%%%%%%%%%%%
\section{Effective moduli parameter energy}
\label{sec:separation}

To further confirm the phase structure associated with the fermionic DWSK, we explore an extension of the moduli parameter--a promotion to an effective moduli parameter--that serves as a measure of separation energy between (two) fermionic DWSks on either side of the CSL. This is important, for if separation energy increases between the fermionic DWSks, there will be mutual repulsion, and it would be associated with a shift of the phase boundary (common with the full DWSk theory) in eq.~\eqref{eq:phase}. And if the separation energy is zero, and a force-free scenario, the phase boundary is as indicated above; we indeed find a force-free scenario, which we now show.

Here, for simplicity, we take the single chiral soliton background
\begin{eqnarray}
    u = e^{i\chi_0}\,\qquad \chi_0 = 4 \arctan e^{m_\pi z}\,.
    %u_0 = e^{i\chi_0}\,\qquad \chi_0 = 4 \arctan e^{m_\pi z}\,.
\end{eqnarray}
Then, the original field $\Sigma$ rotated by SU$(2)$ is given by eq.~\eqref{eq:Sigma}. Now, we simply consider the moduli field $\phi$ of the single anti-lump. That is, we substitute
\begin{eqnarray}
    \phi = \frac{1}{\sqrt{1 + |f|^2}}
    \left(
        \begin{array}{c}
            1\\
            f
        \end{array}
    \right)\,,\qquad
    f = \frac{b}{w^*-d}\,,\qquad (w = x + i y)\,,
\end{eqnarray}
into $\Sigma$, resulting in $\Sigma(x,y,z)$ which is dependent on three spatial coordinates. If we do this with a constant $d$ (the lump's position moduli parameter), we just reproduce the same results that we have already obtained above. Here we add one more trick that we promote $d$ to a function of $z$
\begin{eqnarray}
    d \to d(z)
\end{eqnarray}
that we hereafter refer to as the effective moduli parameter. Let us then determine the effective DW energy of both the ChPT and WZW Hamiltonian, beginning with the former.

\begin{figure}
    \centering
    \includegraphics[width=0.95\linewidth]{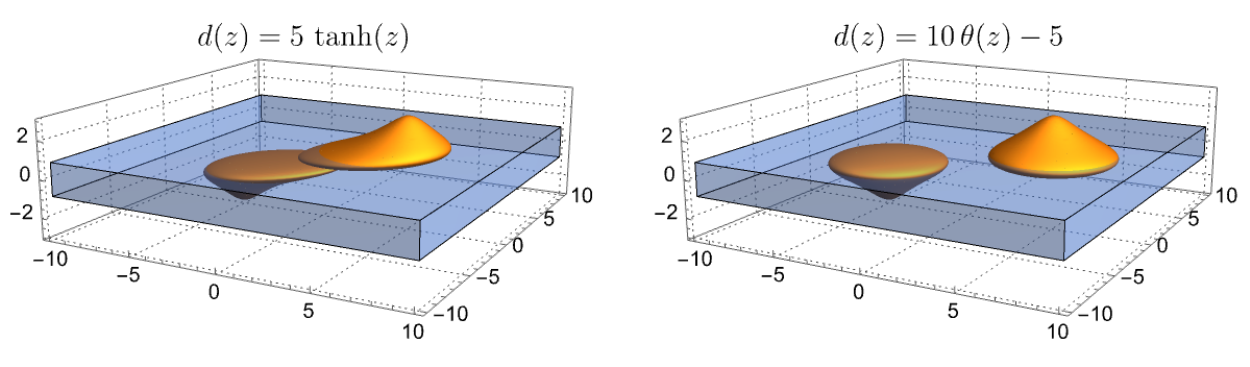}
    \caption{The separation of the macaron into two halves by a smooth step function (left) and by a step function (right).}
    \label{fig:2_half_macaron}
\end{figure}

Now, we plug $\Sigma(x,y,z)$ into the ChPT Hamiltonian. (We will treat the WZW term afterwards.) First, let us take the terms including the derivatives by $x$ and $y$. We find
\begin{eqnarray}
    {\cal H}_{\rm ChPT}^{(x,y)} &=& \frac{16 f_\pi^2 |b|^2 \tanh ^2m_\pi z~ \text{sech}^2m_\pi z}{\left( (x - d(z))^2+y^2+|b|^2\right)^2}\,,\\
    E_{\rm ChPT}^{(x,y)} &=& \int d^3x~{\cal H}_{\rm ChPT}^{(x,y)}
    = \frac{32 \pi  f^2_\pi}{3 m_\pi}\,.
\end{eqnarray}
This corresponds to twice the Skyrmion mass $\pi {\cal C}(1) = 16\pi f_{\pi}^{2}/(3m_{\pi})$ evaluated in the effective theory because we integrated from $-\infty$ to $+\infty$ along the $z$ axis.
Next, we evaluate the rest term of the ChPT Hamiltonian. It can be decomposed into two parts as
\begin{eqnarray}
    {\cal H}_{\rm ChPT}^{(z)} = {\cal H}_{\rm ChPT}^{(z;1)}
    + {\cal H}_{\rm ChPT}^{(z;2)}\,,
\end{eqnarray}
where we have defined
\begin{eqnarray}
    {\cal H}_{\rm ChPT}^{(z;1)} &=& \frac{4 f_\pi^2 m_\pi^2}{\cosh^2m_\pi z}\,,\\
    {\cal H}_{\rm ChPT}^{(z;2)} &=& \frac{8 f_\pi^2 |b|^2 \tanh ^2m_\pi z~ \text{sech}^2m_\pi z}{\left((x - d(z))^2+y^2+|b|^2\right)^2}~d'(z)^2\,.
\end{eqnarray}
Integrating ${\cal H}_{\rm ChPT}^{(z;1)}$ along $z$, we get the soliton tension
\begin{equation}
    %{\cal E}^{\text{(1-soliton)}} 
    \sigma^{\text{(1-soliton)}} = \int dz~{\cal H}_{\rm ChPT}^{(z;1)} = 8 m_\pi f_\pi^2\,.
\end{equation}
On the other hand, by integrating ${\cal H}_{\rm ChPT}^{(z;2)}$ over the $xy$ plane, we get the energy cost of promotion $d \to d(z)$
\begin{eqnarray}
    \int dxdy~{\cal H}_{\rm ChPT}^{(z;2)} = 8 \pi  f_\pi^2 \tanh ^2m_\pi z~ \text{sech}^2m z\, d'(z)^2\,.
\end{eqnarray}
It is positive semidefinite. Therefore, the promotion costs energy in general, but there is an exception:
\begin{eqnarray}
    d(z) = D \theta(z) + d_0\,,
    \label{eq:d_theta_function}
\end{eqnarray}
where $D$ is the separation between two lumps and $d_0$ is the center of mass position of the two lumps.
For this case, the energy cost is zero because $d'(z) \tanh^2 m_\pi z \propto \delta(z) \tanh^2 m_\pi z$ is zero everywhere.  In fig.~\ref{fig:2_half_macaron}, we illustrate how a macaron is separated into two halves by a smooth step function, which is energetically unfavored, and by a step function, which is energetically favorable. If we take the limit \( D \to \infty \) (with \( d_0 \) scaled appropriately), we are left with the half macaron.

Let us also evaluate the WZW term given in eq.~(\ref{eq:WZW_L}). 
The Hamiltonian is 
\begin{equation}
    \mathcal{H}_{\textrm{WZW}}=A_{0}^{B}j_{\textrm{GW}}^{0}
    =-\mu_{\rm B} \frac{\epsilon^{0ijk}}{24\pi^{2}}\mathrm{tr}\bigl(L_{i}L_{j}L_{k}-3ie\partial_{i}[A_{j}Q(L_{k}+R_{k})]\bigr)\,,
\end{equation}
The first term (the baryon number density) is not affected by the promotion $d\to d(z)$. Actually, it reads
\begin{eqnarray}
    {\cal H}_{\rm WZW}^{(1)} = -\mu_{\rm B} \frac{\epsilon^{0ijk}}{24\pi^{2}}\mathrm{tr}\left(L_{i}L_{j}L_{k}\right)
    = - \frac{16 m \mu_{\rm B} |b|^2 \tanh ^2mz \,\text{sech}^3m z}{\pi ^2 \left((x-d(z))^2+y^2+|b|^2\right)^2}\,,
\end{eqnarray}
and the volume integration of it is exactly
\begin{eqnarray}
    \int d^3x\, {\cal H}_{\rm WZW}^{(1)} = -2 \mu_{\rm B}\,.
\end{eqnarray}
Next, we evaluate the second term
\begin{eqnarray}
    {\cal H}_{\rm WZW}^{(2)} = \frac{i e \mu_{\rm B} \epsilon^{0ijk}}{8\pi^{2}}\mathrm{tr}\bigl(\partial_{i}[A_{j}Q(L_{k}+R_{k})]\bigr)\,.
\end{eqnarray}
This is a sum of the following three terms
\begin{eqnarray}
    {\cal H}_{\rm WZW}^{(2;1)} &=& 
    \frac{i e \mu_{\rm B}}{8\pi^{2}}\partial_1 \mathrm{tr}\bigl[A_{2}Q(L_{3}+R_{3}) - A_{3}Q(L_{2}+R_{2})\bigr] \,,
    \label{eq:H2_WZW1}\\
    {\cal H}_{\rm WZW}^{(2;2)} &=& 
    \frac{i e \mu_{\rm B}}{8\pi^{2}}\partial_2 \mathrm{tr}\bigl[A_{3}Q(L_{1}+R_{1}) - A_{1}Q(L_{3}+R_{3})\bigr]\,, 
    \label{eq:H2_WZW2}\\
    {\cal H}_{\rm WZW}^{(2;3)} &=& 
    \frac{i e \mu_{\rm B}}{8\pi^{2}}\partial_3 \mathrm{tr}\bigl[A_{1}Q(L_{2}+R_{2}) - A_{2}Q(L_{1}+R_{1})\bigr]\,.
    \label{eq:H2_WZW3}
\end{eqnarray}
Note that these are the total (spatial) divergence. We first evaluate ${\cal H}_{\rm WZW}^{(2;1)}$ and ${\cal H}_{\rm WZW}^{(2;2)}$.
We have
\begin{eqnarray}
    &&\mathrm{tr}\bigl[A_{2}Q(L_{3}+R_{3}) - A_{3}Q(L_{2}+R_{2})\bigr] = 
    2 i B m \, \text{sech}\,m z \frac{x\left(\rho^2-|b|^2\right)}{\left(\rho^2+|b|^2\right)}\nonumber\\
    && \qquad + 2 i B (\sinh 3 m z-7 \sinh m z) \,\text{sech}^4m z\frac{|b|^2  x (x-d(z))  }{\left(\rho^2 + |b|^2\right)^2}d'(z)\,,
    \label{eq:WZW_2nd_1}\\
    &&\mathrm{tr}\bigl[A_{3}Q(L_{1}+R_{1}) - A_{1}Q(L_{3}+R_{3})\bigr] = 
    2 i B m \, \text{sech}\,m z \frac{y\left(\rho^2-|b|^2\right)}{\left(\rho^2+|b|^2\right)}\nonumber\\
    && \qquad + 2 i B (\sinh 3 m z-7 \sinh m z) \,\text{sech}^4m z\frac{|b|^2  y (x-d(z))  }{\left(\rho^2 + |b|^2\right)^2}d'(z)\,,
    \label{eq:WZW_2nd_2}
\end{eqnarray}
with $\rho^2 = (x-d(z))^2+y^2$. We substitute these into eqs.~(\ref{eq:H2_WZW1}) and (\ref{eq:H2_WZW2}) and integrate them over the $x^1x^2$ plane. However, since the factor $x^i (x-d(z))/(\rho^2 + |b|^2)^2$ ($i=1,2$) rapidly goes to zero as $|x+iy| \to \infty$, the integrations of the second terms of eqs.~(\ref{eq:WZW_2nd_1}) and (\ref{eq:WZW_2nd_2}) vanish. Hence, we can drop the terms proportional to $d'(z)$, and the remaining contributions are 
${\cal H}_{\rm WZW}^{(2;1)} + {\cal H}_{\rm WZW}^{(2;2)}$ can be written as follows:
\begin{eqnarray}
    \label{eq:HWZW1and2}
    {\cal H}_{\rm WZW}^{(2;1)} + {\cal H}_{\rm WZW}^{(2;2)}
    = - \frac{e m \mu_{\rm B}B_z}{2 \pi ^2}
    \left\{1 
    - \frac{2|b|^2 \left(|b|^2-d(x-d)\right)}{\left(|b|^2+(x-d)^2+y^2\right)^2} 
    \right\}\text{sech}\,m z\,.
\end{eqnarray}
Then the volume integral reads
\begin{equation}
    \int d^3x 
    \left({\cal H}_{\rm WZW}^{(2;1)} + {\cal H}_{\rm WZW}^{(2;2)}\right) = 
    - \frac{e\mu_{\rm B}B_z}{2\pi}A + e \mu_{\rm B} B_z |b|^2\,,
\end{equation}
where $A = \int dxdy$ stands for area on the background chiral soliton.
This does not involve $d'(z)$, which implies that the promotion $d \to d(z)$ does not affect at all. 
Similarly, we can evaluate ${\cal H}_{\rm WZW}^{(2;3)}$. We have
\begin{equation}
    \mathrm{tr}\bigl[A_{1}Q(L_{2}+R_{2}) - A_{2}Q(L_{1}+R_{1})\bigr] = 
    -2 i B \sin \left(8 \tan ^{-1}e^{m z}\right)
    \frac{|b|^2 \left(\rho^2+(x-d) d\right)}{\left(\rho^2 + |b|^2\right)^2}
    \,,
    \label{eq:WZW_2nd_3}
\end{equation}
and therefore the volume integral of ${\cal H}_{\rm WZW}^{(2;3)}$ vanishes as
\begin{equation}
    \int {\cal H}_{\rm WZW}^{(2;3)}\,d^3x = \int dxdy
    \left[
    -2 i B \sin \left(8 \tan ^{-1}e^{m z}\right)
    \frac{|b|^2 \left(\rho^2+(x-d) d\right)}{\left(\rho^2 + |b|^2\right)^2}
    \right]^{z \to \infty}_{z \to -\infty} = 0\,.
\end{equation}
In summary, the WZW term gives the following contribution to the mass
\begin{eqnarray}
    \int d^3x~ {\cal H}_{\rm WZW} = 
    -2\mu_{\rm B} - \frac{e\mu_{\rm B}B_z}{2\pi}A + e \mu_{\rm B} B_z |b|^2\,,
\end{eqnarray}
for the single anti-lump, which exactly the same as that with the constant $d$.
Namely, the WZW term is insensitive to the promotion \( d \to d(z) \); therefore, the determination of \( d(z) \) by energy minimization is governed solely by the ChPT term, as already carried out in eq.~(\ref{eq:d_theta_function}).

Combining both the ChPT and WZW terms, we finally find the total mass of the single anti-lump on the single chiral soliton 
\begin{eqnarray}
    M = \left( 8 m_\pi f_\pi^2  - \frac{e\mu_{\rm B}B_z}{2\pi} \right) A + \frac{32 \pi  f^2_\pi}{3 m_\pi}
    -2\mu_{\rm B} + e \mu_{\rm B} B_z |b|^2\,,
\end{eqnarray}
implying that promotion $d \to D\theta(z) + d_0$ does not cost at all.
The first term is proportional to $A$, so it is dominant and gives the condition for the single chiral soliton to appear as a ground state
\begin{eqnarray}
     eB_z \ge \frac{16 \pi m_\pi f_\pi^2}{\mu_{\rm B}}\,.
\end{eqnarray}
The second (subdominant) term is essentially same as eq.~(\ref{eq:energy}) for $k=1$, and gives the condition for the single anti-lump to appear on the single chiral soliton background
\begin{eqnarray}
    \mu_{\rm B} \ge \frac{16 \pi  f^2_\pi}{3 m_\pi}\,.
\end{eqnarray}
Hence, everything is consistent with the results obtained through the effective theory carried out in sec.~\ref{sec:half}.

\section{Chiral limit}
\label{sec:chiral}

Let us last explore the chiral limit of the half DWSk theory to ensure its validity in such regimes; see also ref.~\cite{Amari:2025twm}. The limit occurs for $m_{\pi}\to 0$. Most notably in the chiral limit we observe according to eq.~\eqref{eq:kappa_constraint} that the elliptic modulus governing the Skyrmion density goes to
\begin{equation}
    \kappa\sim\frac{8\pi^{2}m_{\pi}f_{\pi}^{2}}{e\mu_{B}B}=: \frac{m_{\pi}}{\Lambda_\chi}\,.
\end{equation}
Let us first examine the CSL to see how the chiral limit affects its structure.

\begin{figure}
\centering
\includegraphics[scale=0.5]{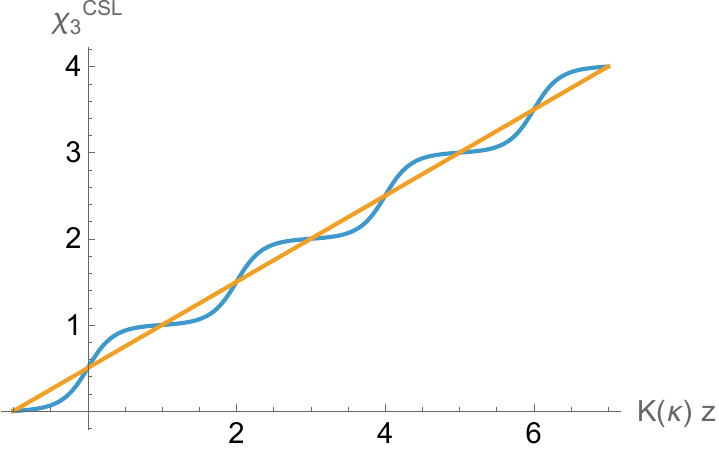}
\caption{
$\chi_3^{\text{CSL}}$ of the CSL shown for both $\kappa=1000/1001$ (blue) given by eq.~\eqref{eq:CSL_solution} and $\kappa\to0$ (yellow) in the chiral limit given by eq.~\eqref{eq:CSL_chiral_limit}. For comparison both are shown against their respective period of $K(\kappa)z$. Notice the flattening linear profile in the chiral limit.
}
\label{fig:csl_chiral}
\end{figure}

 Consider the equation of motion for the CSL in the chiral limit, namely $\partial_{z}^{2}\chi_{3}=0$; we can see that $\chi_{3}(z)$ will simply be a straight line. But we will still have \textit{periodicity}, since $\ell/2=\kappa K(\kappa)/m_{\pi}\sim4\pi^{3}f_{\pi}^{2}/e\mu_{B}B=\pi/2\Lambda_\chi$ will become a constant. Let us see its asymptotic form: 
\begin{equation}
     \chi_{3}^{\textrm{CSL}}=2\mathrm{am}(\Lambda_\chi z,m_{\pi}\Lambda^{-1}_\chi)+\pi=2\Lambda_\chi z+\pi+\mathcal{O}(m_{\pi}^{2})\,.
     \label{eq:CSL_chiral_limit}
\end{equation}
To illustrate the linear structure of the CSL in the chiral limit see fig.~\ref{fig:csl_chiral}, in which both $\kappa=1000/1001$ (blue) and $\kappa\to0$ (yellow) are shown as a function of period.
We remark that if one were to take a naive chiral limit, the width of each domain wall, $m_\pi^{-1}$ would appear to diverge. However, as demonstrated above what happens in the chiral limit is that each domain wall intersects with its nearest neighbor domain wall, blending their periodic structure into the flat linear profile as seen in fig.~\ref{fig:csl_chiral}.

Now let us turn our attention to the case of the shape of the Skyrmion. In the chiral limit the half period Skyrmion charge density, eq.~\eqref{eq:beta}, becomes
\begin{equation}
    \mathcal{B} \sim\frac{16}{\pi}\Lambda\sin^{2}(\Lambda z)\cos^{2}(\Lambda z)q(x,y)+\mathcal{O}(m_{\pi}^{2})\,,
\end{equation}
where we can see that the pion mass has dropped out, and the effective half charge density is no longer of a half macaron type but is spherical. A plot showing the limiting form can be seen below for $\mathcal{B}=(50\pi^{2})^{-1}$ for both halves, with yellow $[0,\pi/2\Lambda_\chi]$ and green $[\pi/2\Lambda_\chi,\pi/\Lambda_\chi]$ along with their period multiples. 
\begin{figure}
\centering
\includegraphics[scale=0.4]{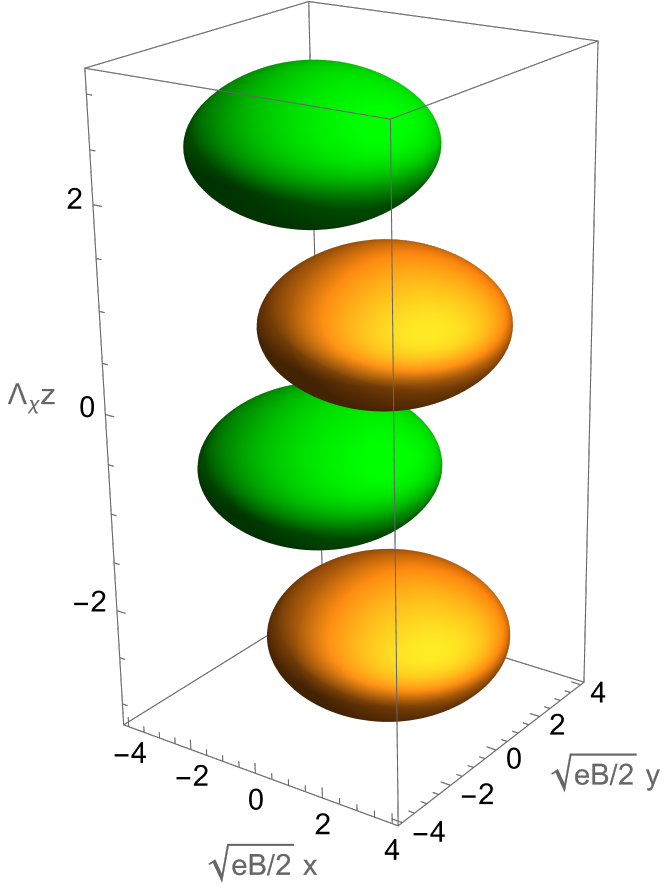}
\caption{Skyrmion charge density isosurface at $\mathcal{B}=(50\pi^{2})^{-1}$ for `half' $n\ell<z<\ell/2+n\ell$ shown in amber and `half' $\ell/2+n\ell<z<(n+1)\ell$ shown in green $\forall n\in \mathbb{Z}$. Both are shown for $\kappa\to 0$ in the chiral limit. In between each Skyrmion sits alternating vacuua at $\Sigma=\pm 1$ (not shown in figure). The baby Skyrmion charge density has been evaluated to eq.~\eqref{eq:baby_Skyrmion} for the single lump with $\omega=x+iy\pm d$ for offset constant $d$.}
\label{fig:beta_chiral}
\end{figure}
The CSL has not been explicitly shown above since the domain walls at each intersection will touch producing the linear profile as described above.

Let us next consider the effective energy in the chiral limit. Since the K\"ahler class goes as
\begin{align}
\mathcal{C}(\kappa) & \sim\frac{8f_{\pi}^{2}}{3m_{\pi}}\frac{\pi}{\kappa^{3}}\Bigl[(2-\kappa^{2})\left(1-\frac{1}{4}\kappa^{2}-\frac{3}{64}\kappa^{4}\right)
+2(\kappa^{2}-1)\left(1+\frac{1}{4}\kappa^{2}+\frac{9}{64}\kappa^{4}\right)\Bigr]
\nonumber\\
 & \sim\frac{\pi f_{\pi}^{2}}{m_{\pi}}\kappa=\frac{8\pi^{3}f_{\pi}^{4}}{e\mu_{B}B}\,,
\end{align}
the kinetic energy becomes constant w.r.t. the zero pion mass. And therefore the effective energy becomes in the chiral limit for the half DWSk theory
\begin{equation}
E_{\textrm{DW}}=\frac{8\pi^{4}f_{\pi}^{4}}{e\mu_{B}B}|k|+\mu_{B}k+\frac{e}{2}\mu_{B}B|b_{k-1}|^{2}\,.
\end{equation}
Consequently the DWSk phase transition occurs at
\begin{equation}
    \mu_{c}=\sqrt{\frac{8\pi^{4}f_{\pi}^{4}}{eB}}\,,
\end{equation}
which also corresponds to a critical magnetic field $eB_{c}\sim8\pi^{4}f_{\pi}^{4}/\mu_{B}^{2}=B_{\textrm{CPC}}/2$, for the previously reported charged pion condensation in ref.~\cite{Brauner:2021sci}. The above analysis indicates that our theory does indeed hold in the chiral limit.

%%%%%%%%%%%%%%%%%%%%%%%%%%%%%%%%%%%%%%%%%%%%%%%%%%%%%%%%%
\section{Conclusions}
\label{sec:conclusions}

\begin{figure}
\centering
\subfloat[]{%
\includegraphics[scale=0.35]{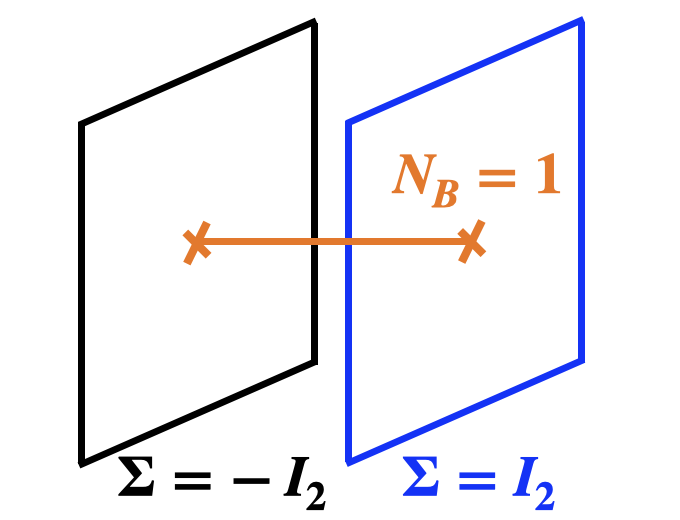}}
\subfloat[]{%
\includegraphics[scale=0.35]{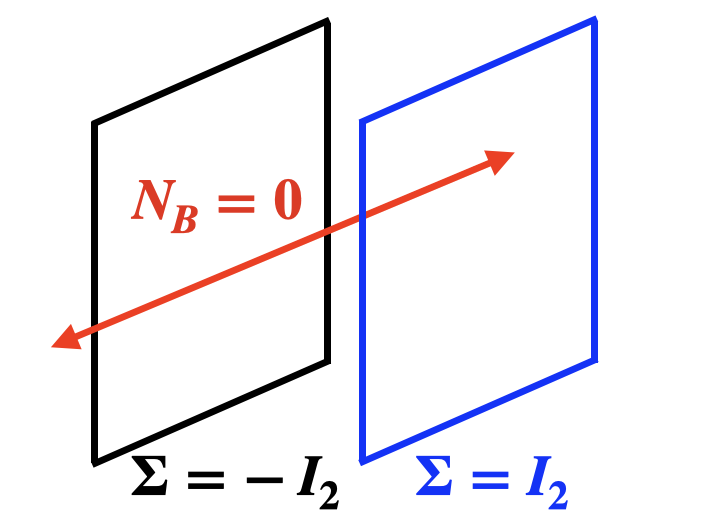}}
\subfloat[]{%
\includegraphics[scale=0.3]{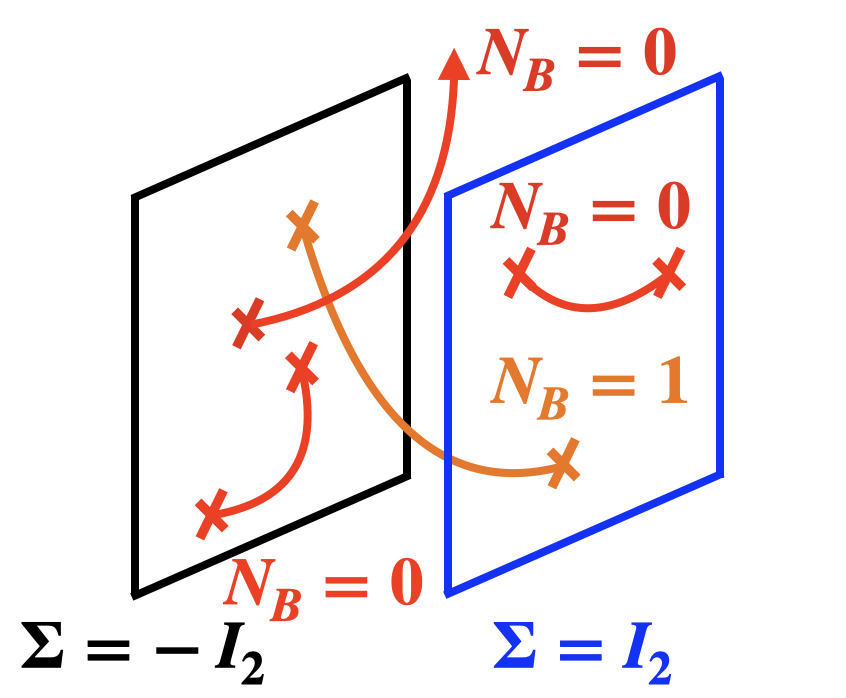}}
\caption{
General configurations of fermionic DWSks 
and topological nature of the baryon number.
The $SU(2)$ flavor symmetry is spontaneously broken to $U(1)$ 
and $\text{SU}(2)_V/\text{U}(1)_3\cong \mathbb{C}P^1\simeq S^2$ 
moduli appear between 
the two planes 
$\Sigma=-\boldsymbol{I}_2$ (black) and $\Sigma=+\boldsymbol{I}_2$ (blue), 
while on the boundary planes  
the $SU(2)$ is recovered.
(a) 
A fermionic DWSk is described by an open string (orange line) interpolating between the two with baryon topological charge density $N_B=1$. 
(b) An open string parallel to the planes carries no baryon number. 
(c) Here, a DWSk's interpolating path has been changed due to the effective moduli parameter promotion. Notice that paths in which the two planes are not connected give zero baryon topological charge (in red).
}
\label{fig:vacua}
\end{figure}

Previous studies of DWSks of ChPT in QCD in a strong magnetic field~\cite{Eto:2025fkt,Eto:2023wul} have hinted at the separation of the minimal Skyrmion with $N_B=2$ characterized with bosonic statistics \cite{Amari:2024mip} into a true minimal Skyrmion theory with \textit{fermionic} $N_B=1$. In this work we have established that this is indeed the case. On the CSL with period $\ell=2\kappa K(\kappa)/m_\mu$ the ground state too has periodic structure winding of the pion field for example from $\Sigma=-\boldsymbol{I}_2$ to $+\boldsymbol{I}_2$ to $-\boldsymbol{I}_2$ in SU$(2)_V$. On half of the period, from e.g. 0 to $\ell/2$, however, we have found the existence of DWSks that interpolate from $\Sigma=-\boldsymbol{I}_2$ to $+\boldsymbol{I}_2$, whose effective theory admits lumps that cover $\pi^2(\mathbb{C}P^1)$. The minimal DWSk for the single soliton case, and multiple soliton cases with $\kappa\lesssim 1$, has isosurface of half macaron shape, and moreover possesses fermionic spin-statistics. This was confirmed using the Witten approach~\cite{Witten:1983tx,Amari:2024mip}. We further found by introducing an effective moduli parameter that there was no energy cost associated with separating the two fermionic DWSks attached on either side of the CSL. The promotion of the DWSk onto an effective moduli parameter is also represented in Fig.~\ref{fig:vacua}. The zero separation energy confirmation was also important to establish the phase structure as previously found for the full theory in refs.~\cite{Eto:2025fkt,Eto:2023wul}. Indeed with repulsion between the DWSks on either side one would find the phase boundary between the DWSk and CSL phases to be shifted; we found the scenario of no force however. The chiral limit of the fermionic DWSk of QCD in a magnetic field was also examined in which the CSL reduces to one with linear dependence, and the DWSks to take on a symmetric shape. But otherwise we found our fermionic DWSk theory to be robust even in the chiral limit. 

The scaling instability of DWSks can be cured by either higher-derivatives~\cite{Eto:2025fkt} or gauge field dynamics~\cite{Amari:2024adu,Amari:2024fbo}, which were studied for bosonic DWSks. This could be generalized to the fermionic DWSk studied in this paper.

As we have shown the DWSk of the full theory may be formulated into a theory of  two species of fermions, it is intriguing for further studies of phase structure. 
For instance, the effective theory on the chiral soliton is conjectured to be characterized by a symmetry-protected topological order. 
Description by two species of fermions may be useful for such a study.

In this paper, we have studied fermionic DWSks
based on the effective theory on a half period of the CSL.
However, apart from the effective description 
in 2+1 dimensions, 
we can consider more general configurations from a 3+1 dimensional bulk point of view. 
In such a case, the $SU(2)$ flavor symmetry 
is spontaneously broken to $U(1)$ almost everywhere 
except for 
$\Sigma = - \boldsymbol{I}_2$ 
(denoted by the black planes in  
fig.~\ref{fig:vacua}) and 
$\Sigma = + \boldsymbol{I}_2$ 
(denoted by the blue planes in  
fig.~\ref{fig:vacua}).
They 
correspond to the centers of 
the chiral soliton solitons, 
and the vacuum, respectively, in the CSL, 
and appear alternately along 
the $z$ direction. 
Therefore, in a region between 
$\Sigma = - \boldsymbol{I}_2$ and 
$\Sigma = + \boldsymbol{I}_2$, 
there can exist
topologically stable 
lump strings supported by 
 the topological charge  $\pi_2({\mathbb C}P^1) \simeq {\mathbb Z}$.
First, the configurations studied in this paper are independent of the $z$ coordinate  and can be represented 
in 
fig.~\ref{fig:vacua}(a) 
as an open string (the orange line)
stretched between the two planes. 
While the lump strings have a topological charge, they can end on 
$\Sigma = - \boldsymbol{I}_2$ or 
$\Sigma = + \boldsymbol{I}_2$, where 
the $SU(2)$ flavor symmetry is recovered 
and the strings can be cut due to 
$\pi_2 [SU(2)]=0$. 
Interestingly the configuration looks like 
an open string stretched between two D-branes in string theory.
Fig.~\ref{fig:vacua}(b) shows 
a infinitely long lump string perpendicular to the 
$z$ coordinate, carrying no baryon number.
As illustrated in fig.~\ref{fig:vacua}(c),
only when an open lump string connects 
$\Sigma = - \boldsymbol{I}_2$ and
$\Sigma = + \boldsymbol{I}_2$, it carries 
a baryon number one $N_{\rm B}=1$; 
an open string only one of whose 
ends terminates on the plane 
and one whose both ends terminate 
on the same plane 
carry no baryon number. 
These give a topological nature of the baryon number.

DWSks (for the full theory--and we may immediately infer here as well for the fermionic DWSks) are charged once the effects of a dynamical gauge coupling are incorporated~\cite{Amari:2024fbo}. Then it is an interesting question to ask about the baryon density on the CSL, and its possible implications to the phase diagram, since a Debye length between Skyrmions would be present both in equilibrium and at finite temperature. This is a subject of future work.

\section*{Acknowledgments}
%\comPC{Can you all please fill out your JSPS grant info? Thanks!} 

This work is supported in part by JSPS Grant-in-Aid for Scientific Research KAKENHI Grant No. JP22H01221 and JP23K22492 (M.~E., M.~N. and Z.~Q.). 
This work is also supported in part by the WPI program ``Sustainability with Knotted Chiral Meta Matter (WPI-SKCM$^2$)'' at Hiroshima University. P.C. would further like to acknowledge support from the Research Start-up Support Fund of WPI-SKCM$^2$.

\bibliographystyle{JHEP}
\bibliography{references}
\end{document}